\input{aipcheck}

\documentclass[
    ,final            
  ]
  {aipproc}

\layoutstyle{6x9}

\begin{document}

\title{QCD Vacuum Topology and Glueballs}

\author{Hilmar Forkel}{
  address={IFT - Universidade Estadual Paulista, Rua Pamplona, 145, 01405-900 Sao Paulo,
SP, Brazil, and Institut f\"{u}r Theoretische Physik, Universit\"{a}t Heidelberg, D-69120
Heidelberg, Germany}
}

\begin{abstract}
We outline a comprehensive study of spin-0 glueball properties which, in
particular, keeps track of the topological gluon structure. Specifically, we
implement (semi-hard) topological instanton physics as well as topological
charge screening in the QCD vacuum into the operator product expansion (OPE)
of the glueball correlators. A realistic instanton size distribution and 
the (gauge-invariant) renormalization of the instanton contributions are 
also implemented. Predictions for $0^{++}$ and $0^{-+}$
glueball properties are presented.
\end{abstract}

\maketitle


\section{Introduction}

While the naive quark-model description of the classical hadrons is
surprisingly successful, we lack an equally simple and systematic approach to
hadrons which are not predominantly built from valence quarks. Glueballs
\cite{gel72}, containing no valence quarks at all, are a case in point:
although ``valence gluons'' had a limited amount of success in glueball
models, their role as relevant degrees of freedom (as opposed to strings,
e.g.) is much less clear. This is partly because, in contrast to valence
quarks, gluons cannot be traced by observable internal quantum numbers. In the
following, we sketch a recent study of glueballs which instead relies on some
robust topological gluon properties to keep track of the elusive hadronic
glue. A comprehensive analysis of both spin-0 glueball channels along these
lines can be found in Ref. \cite{for03}.

\section{Sketch of approach}

Our approach is based on the correlations functions of the scalar $\left(
0^{++}\right)  $ and pseudoscalar $\left(  0^{-+}\right)  $ glueball
channels,
\begin{equation}
\Pi_{G}(-q^{2})=i\int d^{4}x\,e^{iqx}\left\langle 0|T\,O_{G}\left(  x\right)
O_{G}\left(  0\right)  |0\right\rangle
\end{equation}
where $O_{G}$ with $G\in\left\{  S,P\right\}  $ are the standard gluonic
interpolating fields (with lowest mass dimension): $O_{S}\left(  x\right)  
=\alpha_{s}G_{\mu\nu}^{a}\left(  x\right) G^{a\mu\nu}\left(  x\right)$ and 
$O_{P}\left(  x\right)   =\alpha_{s}G_{\mu\nu}^{a}\left(  x\right) 
\tilde{G}^{a\mu\nu}\left(  x\right)$. The zero-momentum limits of these 
correlators are governed by low-energy
theorems (LETs) \cite{let} which impose stringent
constraints on the sum rules and require nonperturbative
short-distance physics \cite{for01,for03}.

Our theoretical framework for calculating the glueball correlators at short
distances, i.e. large, spacelike momenta $Q^{2}\equiv-q^{2}\gg\Lambda_{QCD}$, 
is the instanton-improved operator product expansion (IOPE)
\begin{equation}
\Pi_{G}(Q^{2})=\sum_{D=0,4,...}\tilde{C}_{D}^{\left(  G\right)  }\left(
Q^{2}; \mu\right)  \left\langle \hat{O}_{D}\right\rangle_{\mu} ,
\end{equation}
which factorizes ``hard'' mode (with momenta $\left|  k\right| 
 >\mu$) and ``soft'' mode contributions (with $\left|  k\right|  \leq\mu$). 
The perturbative Wilson coefficients are standard. In Ref. \cite{for03} 
we calculate and study nonperturbative contributions due to direct 
instantons \cite{inst} and topological-charge screening. 

In order to make contact with the hadronic information contained in the
glueball correlators, one writes the dispersive representation%
\begin{equation}
\Pi_{G}\left(  Q^{2}\right)  =\frac{1}{\pi}\int_{0}^{\infty}%
ds\frac{Im \Pi_{G}\left(  -s\right)  }{s+Q^{2}} 
\label{disprel}
\end{equation}
(necessary subtractions implied) and inserts the standard sum-rule 
description of the spectral functions,
\begin{equation}
Im \Pi_{G}^{\left(  ph\right)  }\left(  s\right)
= Im\Pi_{G}^{\left(  pole\right)  }\left(  s\right)
+ Im \Pi_{G}^{\left(  cont\right)  }\left(  s\right),
\label{phspecdens}%
\end{equation}
which contains one or two resonance poles $ Im \Pi^{\left(  
pole\right)  }\left(  s\right)   =\pi
\sum_{i=1}^{2}f_{Gi}^{2}m_{Gi}^{4}\delta\left(  s-m_{Gi}^{2}\right)$ 
and the local-duality continuum 
$ Im \Pi_{G}^{\left(  cont\right)  }\left(  s\right) 
=\theta\left(  s-s_{0}\right)  Im \Pi_{G}^{\left(  
IOPE\right)}\left(  s\right)$.

One then writes QCD sum rules by matching the Borel-transformed
IOPE correlators - weighted with powers of $-Q^{2}$ - to these  
phenomenological counterparts, whose glueball parameters are 
thereby determined. 

\section{Results and conclusions}

Our study  \cite{for03} 
has produced several remarkable and
phenomenologically important results. Contrary to naive
expectation, much of the topological gluon physics was found to happen at
surprisingly short distances $\left|  x\right|  \sim0.2-0.3$ fm. To study 
its impact on glueball properties, we have performed a
comprehensive quantitative analysis of eight Borel sum rules. 
The results reveal a rather diverse pattern of spin-0 glueball
properties. 

In the scalar channel, the improved treatment of the direct-instanton sector
reduces our earlier (spike-distribution based) result for the $0^{++}$
glueball mass by about 20\%, to $m_{S}=1.25\pm0.2$ GeV. This is somewhat
smaller than the quenched lattice results \cite{lat} which are, however,
expected to be reduced by light-quark effects and quarkonium admixtures. Our
mass prediction is consistent with the broad glueball state found in a recent
$K$-matrix analysis \cite{ani03} (including the new Crystal Barrel states).
The systematics in our results from different Borel-moment sum rules indicates
a rather large width of the scalar glueball, $\Gamma_{S}\sim 0.3$ GeV. Our
prediction for the glueball decay constant, \ $f_{S}=1.05\pm0.1$ GeV, is
several times larger than the value obtained when ignoring the nonperturbative
Wilson coefficients. The implied, small glueball size is 
in agreement with lattice results. Furthermore, our prediction for
$f_{S}$ implies substantially larger partial widths of radiative $J/\psi$ and
$\Upsilon$ decays into scalar glueballs and is therefore important for
experimental glueball searches, in particular for the interpretation of the
recent CLEO \cite{cleo} and forthcoming CLEO-III data on $\Upsilon
\rightarrow\gamma f_{0}$ and other decay branches. 

In the $0^{-+}$ glueball channel we have found compelling evidence for
topological charge screening to provide critical contributions to the IOPE.
Besides restoring the axial Ward identity, they are found to resolve the 
positivity-bound violation and to create
a strong signal for both the $\eta^{\prime}$ and the pseudoscalar glueball. 

After including the screening contributions, all four $0^{-+}$ Borel-moment
sum rules are stable and yield consistent results. (Previous analyses of the
$0^{-+}$ sum rules had discarded the lowest-moment sum rule and therefore missed the
chance to implement the first-principle information from the low-energy
theorem, as well as a very useful consistency check.) The two-resonance fit is
clearly favored over the one-pole approximation, i.e. the IOPE provides clear
signals for the $\eta^{\prime}$ resonance (with small $\eta$ admixtures due to
mixing) in addition to a considerably heavier $0^{-+}$ glueball. Our mass
prediction $m_{P}=2.2\pm1.5$ GeV for the pseudoscalar glueball lies inside the
range obtained from quenched and unquenched lattice data. The coupling
$f_{P}=0.6\pm0.2$ GeV is again enhanced by the nonperturbative Wilson
coefficients, but less strongly than in the scalar channel. The consequently
larger partial width of radiative quarkonium decays into pseudoscalar
glueballs and the enhanced $\gamma\gamma\rightarrow G_{P}\pi^{0}$ cross
section at high momentum transfers will be relevant for the experimental
identification of the lowest-lying $0^{-+}$ glueball and help in measuring 
its properties.

Our nonperturbative IOPE, including the topological short-distance physics,
will be useful for calculating other spin-0 glueball properties as well.
Quantitative estimates of the mentioned production rates in gluon-rich
channels (including $J/\psi$ and $\Upsilon$ decays) and characteristic
glueball decay properties and signatures, including $\gamma\gamma$ couplings,
OZI suppression and branching fractions incompatible with $q\bar{q}$ decay,
would be particularly interesting.

This work was supported by FAPESP.





\bibliographystyle{aipprocl} 


\end{document}